\documentclass[%
 reprint,
superscriptaddress,
 amsmath,amssymb,
]{revtex4-2}

\usepackage{graphicx}
\usepackage{dcolumn}
\usepackage{verbatim} 
\usepackage{bm}
\usepackage[table]{xcolor}
\definecolor{red}{rgb}{1,0,0}
 
\usepackage{comment}

\usepackage{textcomp}

\usepackage{hyperref}
\hypersetup{
colorlinks=true,
linkcolor=black,
filecolor=black,  
citecolor=black,
urlcolor=black,
}

\usepackage{pdfpages} 
\usepackage{pgffor} 

\makeatletter
\AtBeginDocument{\let\LS@rot\@undefined}
\makeatother

\def\supplementfilename{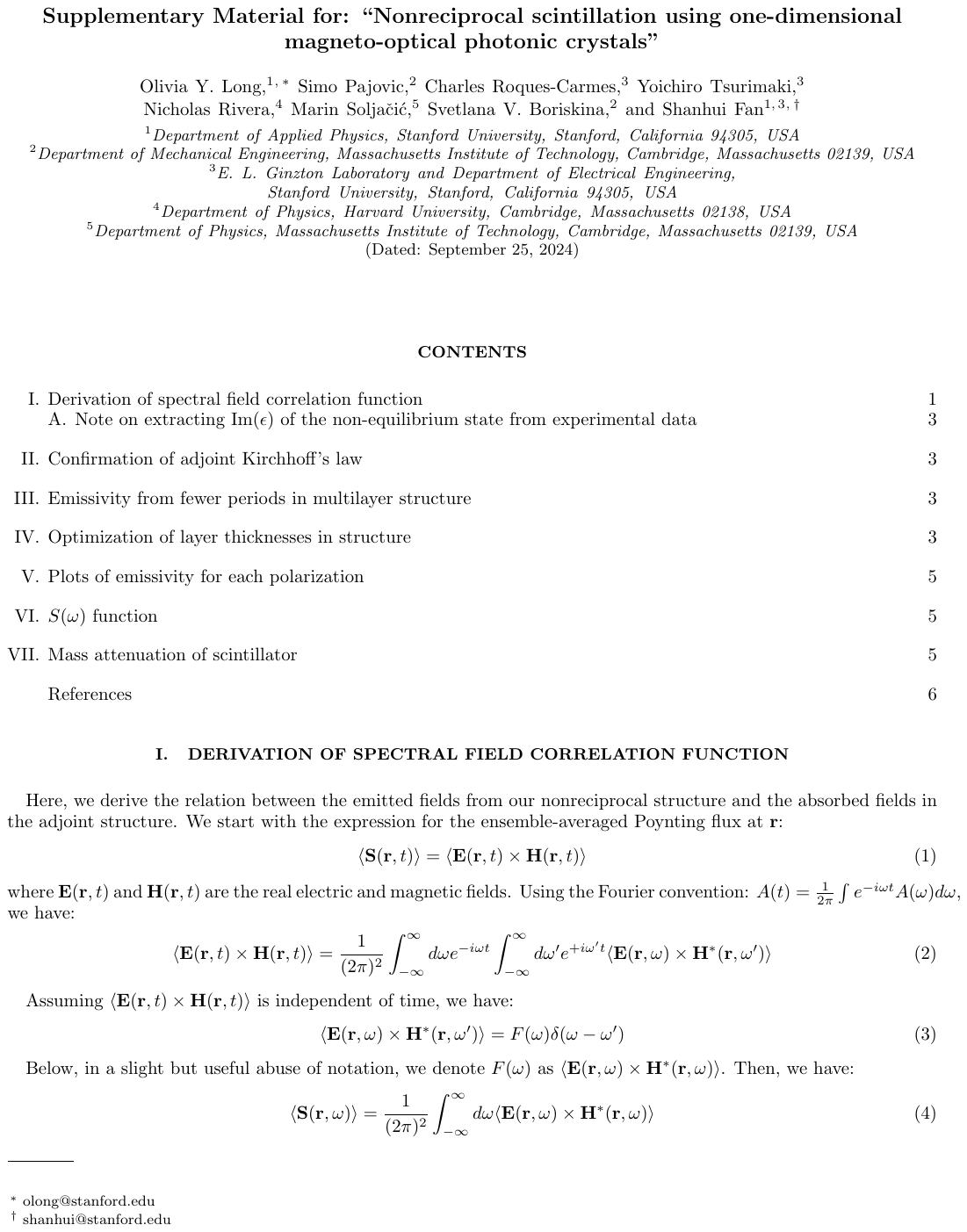}

\pdfximage{\supplementfilename}
\def\numbersupplementpages{\the\pdflastximagepages}

\newif\ifarXiv
\arXivtrue 

\begin{document}


\title{Nonreciprocal scintillation using one-dimensional magneto-optical photonic crystals}

\author{Olivia Y. Long}
\email{olong@stanford.edu}
 \affiliation{Department of Applied Physics, Stanford University, Stanford, California 94305, USA}

\author{Simo Pajovic}
 
\affiliation{Department of Mechanical Engineering, Massachusetts Institute of Technology, Cambridge,
Massachusetts 02139, USA
}%

\author{Charles Roques-Carmes}%
 \affiliation{E. L. Ginzton Laboratory and Department of Electrical Engineering,
Stanford University, Stanford, California 94305, USA}

\author{Yoichiro~Tsurimaki}%
 \affiliation{E. L. Ginzton Laboratory and Department of Electrical Engineering,
Stanford University, Stanford, California 94305, USA}

\author{Nicholas Rivera}
\affiliation{Department of Physics, Harvard University, Cambridge, Massachusetts 02138, USA}

\author{Marin Solja\v{c}i\'{c}}
\affiliation{Department of Physics, Massachusetts Institute of Technology, Cambridge,
Massachusetts 02139, USA
}

\author{Svetlana V. Boriskina}
\affiliation{Department of Mechanical Engineering, Massachusetts Institute of Technology, Cambridge,
Massachusetts 02139, USA
}

\author{Shanhui Fan}

\email{shanhui@stanford.edu}
\affiliation{Department of Applied Physics, Stanford University, Stanford, California 94305, USA}
\affiliation{E. L. Ginzton Laboratory and Department of Electrical Engineering,
Stanford University, Stanford, California 94305, USA}

\date{\today}

\begin{abstract}
Scintillation describes the conversion of high-energy particles into light in transparent media and finds diverse applications such as high-energy particle detection and industrial and medical imaging. This process operates on multiple timescales, with the final radiative step consisting of spontaneous emission, which can be modeled within the framework of quasi-equilibrium fluctuational electrodynamics. Scintillation can therefore be controlled and enhanced via nanophotonic effects, which has been proposed and experimentally demonstrated.
Such designs have thus far obeyed Lorentz reciprocity, meaning there is a direct equivalence between scintillation emission and absorption by the scintillator. 
However, scintillators that do not obey Lorentz reciprocity have not been explored, even though they represent a novel platform for probing emission which is both nonequilibrium \textit{and} nonreciprocal in nature. In this work, 
we propose to harness nonreciprocity to achieve directional control of scintillation emission, granting an additional degree of control over scintillation.
Such directionality of light output is important in improving collection efficiencies along the directions where detectors are located.
%
%
We present the design of a nonreciprocal scintillator using a one-dimensional magnetophotonic crystal in the Voigt configuration. Our work demonstrates the potential of controlling nonequilibrium emission such as scintillation by breaking reciprocity and expands the space of nanophotonic design for achieving such control. 

\end{abstract}

\maketitle


\section{Introduction}

Scintillation is a prevalent physical phenomenon that finds applications in high-energy particle detection, medical imaging, industrial flaw detection, high resolution two-dimensional imaging, and radio astronomy \cite{inorganic_scint_text_2017, scint_detectors_for_x-rays_Nikl_2006}. 
It involves the conversion of incident high-energy particles such as free electrons, X-rays, and $\gamma$-rays into visible light, often to be detected by photodetectors. 
%
%
Scintillation is characterized by a complex sequence of processes, each described by different time scales: the conversion of incident particle energy into electron-hole pairs ($\sim$ 1 ps), the thermalization of electrons and holes, the transport of electrons and holes through the scintillating material ($\sim$ 1 ps $-$ 10 ns), and luminescence from radiative recombination at a luminescence center ($\sim$ 1 ns) \cite{scint_detectors_for_x-rays_Nikl_2006}.
Since thermalization occurs on a much faster time scale ($\sim$ 1 ps) than spontaneous emission ($\sim$ 1 ns), 
scintillation can be described by emission from a quasi-equilibrium distribution of fluctuating currents ~\cite{light_emission_nonequi_bodies_greffet_PRX_2018, viktar_tutorial_EM_nonreciprocity_2020,framework_scintillation_science_2022}. This means scintillation can be understood in analogy to thermal radiation \cite{rytov1989_vol3}.

Since scintillation is a form of spontaneous emission, recent works have proposed and demonstrated the capabilities of nanophotonic design in enhancing emission rates and imaging resolution \cite{ido_scintillator_PRL_2020, framework_scintillation_science_2022, enhanced_imaging_inv_design_nanophot_scint_kaminer_2023, improving_light_output_CTR_scint_crystals_phc_slabs_2019}, paralleling similar achievements in the design of thermal emitters and absorbers \cite{fan2017thermal}. Such designs integrate scintillators into nanophotonic structures, yielding so-called ``nanophotonic scintillators.'' Previous work relied on two different methods: (1) exploiting the Purcell effect by engineering the local density of photonic states to increase a scintillator's emission rates  \cite{Ye_purcell_scintillator_2024, framework_scintillation_science_2022, purcell_x-ray_scint_arxiv_2023, towards_second_gen_metascint_using_purcell_arxiv_2024}; and (2) employing surface patterns to enhance scintillation outcoupling to a photodetector~\cite{framework_scintillation_science_2022, Lecoq2020}. 
However, all previous works have thus far assumed nanophotonic structures that obey Lorentz reciprocity \cite{framework_scintillation_science_2022, ido_scintillator_PRL_2020}. 

Reciprocity represents a fundamental symmetry of electromagnetism, and is applicable to any linear, time-invariant material system described by symmetric permittivity and permeability tensors \cite{landau_lifshitz_vol8}. In these systems, reciprocity implies Kirchhoff's law of radiation, which states the equality of spectral directional emissivity and absorptivity. Kirchhoff's law has been used for the design of scintillators \cite{framework_scintillation_science_2022}. The consequences of lifting the assumption of Lorentz reciprocity for scintillation -- and in turn, applications of nonreciprocal scintillation for medical imaging and particle detection -- have not yet been explored.

Methods of achieving nonreciprocity include introducing a magneto-optic material, using nonlinear media, or applying time modulation \cite{viktar_tutorial_EM_nonreciprocity_2020}.
Nonreciprocity in electromagnetics has been widely harnessed for technologies such as optical and microwave isolators, circulators, and unidirectional waveguides 
\cite{tutorial_em_nonreciprocity_viktar_2020, electromag_nonrecip_alu_prapplied_2018, nonrecip_therm_photonics_review_2024}. 
More recently, nonreciprocal structures have found notable applications in thermal radiation due to the breakdown of Kirchhoff's law of radiation, that is, unequal emission and absorption along a particular direction for a given frequency and polarization \cite{linxiao_near-complete_violation_PRB_2014, bo_zhao_near-complete_violation_kirchhoff_opt_letters_2019, violating_kirchhoff_law_thermal_rad_yubin_2021, direct_observ_violation_kirchhoff_2023, nonrecip_absorption_ENZ_cwqiu_2023, gold_GAGA_for_magnetophotonic_crystals_nanophotonics_2024, broadband_nonrecip_thermal_atwater_2024}. Such structures have been theoretically shown to be important for achieving the ultimate efficiency limit in solar energy harvesting \cite{yubin_reaching_ultimate_efficiency_solar_energy_nonrecip_2022, ries_1983, martin_green_time-asymm_PV_2012}. The efficiency benefits of nonreciprocity in thermal systems suggest that there may be analogous benefits in systems that can be modeled as being in quasi-equilibrium such as scintillation, with potential applications to imaging and particle detection. 


In this work, we propose the design of a one-dimensional magnetophotonic crystal to achieve nonreciprocal scintillation. In order to design and understand the nonreciprocal scintillator, we develop a framework of fluctuational electrodynamics for a nonreciprocal, quasi-equilibrium system and apply it to model nonreciprocal scintillation emission. 
This theoretical formalism contributes to our overall understanding of emission in non-equilibrium systems with the constraint of reciprocity lifted, paving the way for further exploration in systems such as LEDs. By harnessing nonreciprocity, we are able to achieve off-normal directional emission, which is desirable in detection schemes where photodetectors are located off-normal from the scintillator \cite{scint_counting_birks_textbook_1964}, yielding higher collection efficiency and peak emission intensities along a given angular emission channel. 
We emphasize that this type of asymmetric emission pattern could \textit{not} be achieved with any bulk scintillator or planar reciprocal nanophotonic scintillator alone and is therefore a unique signature of nonreciprocal scintillation. Moreover, the study of scintillation provides a system to explore the consequence of nonreciprocity in a non-equilibrium system. This is in contrast with the study of nonreciprocal thermal emission, where the emitters are usually assumed to be in local thermal equilibrium.  

This paper is organized as follows. Section II covers the theoretical basis for our model of nonreciprocal scintillation. Section IIA describes the scattering matrix analysis for identifying nonreciprocity in our system. In Section IIB, we derive the Green's function relations for nonreciprocal systems. Section IIC describes how fluctuational electrodynamics is employed to model our quasi-equilibrium system and Section IID describes the application of nonreciprocity in our system for directional scintillation. In Section III, we present our proposed magneto-optical photonic crystal structure and illustrate its distinct nonreciprocal scintillation properties. We conclude with discussion and summary in Sections IV and V.

\section{Theory}
\subsection{Scattering matrix analysis}

\begin{figure}
\includegraphics[width=0.4\textwidth]{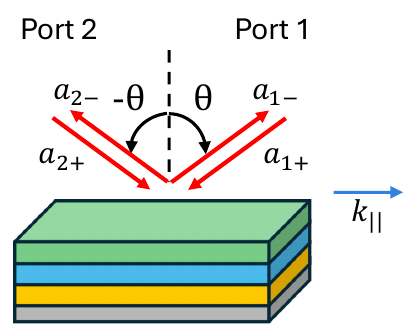}
\caption{\label{s_matrix_fig} Schematic of a two-port system describing plane waves interacting with a dielectric multilayer structure on top of a mirror (gray). The parallel wavevector $k_{\parallel}$ is conserved due to translational symmetry.}
\end{figure}

In the following, we consider a linear, two-port system describing plane waves interacting with a dielectric multilayer structure having a perfect mirror on the back facet. 
Ports 1 and 2 correspond to plane waves propagating in the directions $\theta$ and $-\theta$ from normal, respectively, as shown in Fig. \ref{s_matrix_fig}.
Such an optical system can be described by a scattering matrix $S$ that relates the input wave amplitudes ($a_{1+}, a_{2+}$) to the output amplitudes ($a_{1-}, a_{2-}$) in the ports \cite{kerns1981plane}:
\begin{align}
    S = \begin{pmatrix}
        s_{11} & s_{12} \\
        s_{21} & s_{22}
    \end{pmatrix}
\end{align}
where $(a_{1-}, a_{2-})^T = S (a_{1+}, a_{2+})^T$.
In this system, $k_{\parallel}$ is conserved due to translational symmetry and there is only specular reflection. 
Thus, $s_{11}= s_{22} = 0$ and we have the following $S$-matrix:
\begin{align}
    S = \begin{pmatrix}
        0 & s_{12} \\
        s_{21} & 0 
    \end{pmatrix}
\end{align}

For a system in thermal equilibrium, the emissivity $\varepsilon_i$ into port $i$ can be expressed as: $\varepsilon_i = 1-\sum_j |s_{ij}|^2 $ where $j\in \{1,2\}$. Similarly, the absorptivity $\alpha_i$ in port $i$ is: $\alpha_i = 1- \sum_j |s_{ji}|^2$ \cite{david_miller_universal_radiation_laws_PNAS_2017, nonrecip_therm_photonics_review_2024}. In reciprocal systems, the scattering matrix is symmetric, i.e. $S = S^T$ and thus, $\varepsilon_i = \alpha_i$ \cite{viktar_tutorial_EM_nonreciprocity_2020}. This is the statement of the original Kirchhoff's law of radiation, which assumes reciprocity \cite{kirchhoff_OG_paper}.

However, if the system is nonreciprocal, we may have $|s_{12}| \neq |s_{21}|$.
%
%
In this case, the emissivities in the directions $\theta$ and $-\theta$ are not equal, since $\varepsilon_1 = 1-|s_{12}|^2$ and $ \varepsilon_2 = 1-|s_{21}|^2$.
The contrast $\varepsilon_1 - \varepsilon_2 \neq 0$ is thus a signature of nonreciprocity. Such a signature has recently been experimentally detected to demonstrate nonreciprocity in a thermal emitter \cite{direct_observ_violation_kirchhoff_2023}. Here, we would like to explore such a signature in scintillation, which requires us to develop a fluctuational electrodynamics formalism to describe such nonreciprocal, quasi-equilibrium emission.
%
%

\subsection{Generalized reciprocity}
In this section, we briefly derive the dyadic Green's function relations in nonreciprocal systems and show how the scintillation power can be related to field concentrations in the structure. 

The adjoint structure, or complementary medium, to a structure with permittivity $\bar{\bar{\epsilon}}$, permeability $\bar{\bar{\mu}}$, and no magnetoelectric coupling is defined to have parameters \cite{kong_EM_wave_theory_2008}: 
\begin{align}
\tilde{\bar{\bar{\epsilon}}} &= \bar{\bar{\epsilon}}^T \\
    \tilde{\bar{\bar{\mu}}} &= \bar{\bar{\mu}}^T
\end{align}
where we use the tilde to indicate parameters of the adjoint structure.
For nonmagnetic ($\mu = \mu_0$), magneto-optic materials, the adjoint structure is equivalent to reversing the direction of the external applied magnetic field.
It can be shown that the following generalized reciprocity theorem holds \cite{kong_EM_wave_theory_2008}:
\begin{align}\label{reaction_eqn}
    \int_{V_b} \mathbf{J}_b \cdot \mathbf{E}_a dV_b = \int_{V_a} \mathbf{J}_a \cdot \tilde{\mathbf{E}}_b dV_a
\end{align} 
where $\mathbf{J}_a, \mathbf{J}_b$ are the current sources $a,b$ and $\mathbf{E}_a, \tilde{\mathbf{E}}_b$ are the fields produced by source $a$ in the medium and by source $b$ in its adjoint, respectively.

The electric fields are related to the Green's function through: \begin{align}\label{Efield_greens}
    \mathbf{E}(\mathbf{r}, \omega) = i\mu_0 \omega \int_{V^\prime} d\mathbf{r}^{\prime} \bar{\bar{{G}}}(\mathbf{r}, \mathbf{r}^{\prime}, \omega) \mathbf{J}(\mathbf{r}^{\prime}, \omega )
\end{align}
%
%
Using this relation in Eq. \ref{reaction_eqn}, we have: 
\begin{align}
    \mathbf{J}_b(\mathbf{r}) \cdot \big[ \bar{\bar{{G}}}(\mathbf{r}, \mathbf{r}^{\prime}) \mathbf{J}_a(\mathbf{r}^{\prime} ) \big] &= \mathbf{J}_a(\mathbf{r}^{\prime}) \cdot \big[ \tilde{\bar{\bar{{G}}}}(\mathbf{r}^{\prime}, \mathbf{r}) \mathbf{J}_b(\mathbf{r} ) \big] \nonumber \\
    \mathbf{J}_b^T (\mathbf{r}) \big[ \bar{\bar{{G}}}(\mathbf{r}, \mathbf{r}^{\prime}) \mathbf{J}_a(\mathbf{r}^{\prime} ) \big] &=
    \big[ \tilde{\bar{\bar{{G}}}}(\mathbf{r}^{\prime}, \mathbf{r}) \mathbf{J}_b(\mathbf{r} ) \big]^T \mathbf{J}_a(\mathbf{r}^{\prime}) \nonumber \\
    \mathbf{J}_b^T (\mathbf{r})  \bar{\bar{{G}}}(\mathbf{r}, \mathbf{r}^{\prime}) \mathbf{J}_a(\mathbf{r}^{\prime} ) &= \mathbf{J}_b^T(\mathbf{r} ) \tilde{\bar{\bar{{G}}}}^T(\mathbf{r}^{\prime}, \mathbf{r}) \mathbf{J}_a(\mathbf{r}^{\prime})
\end{align}
where $\tilde{\bar{\bar{G}}}$ is the Green's function in the adjoint structure.

Thus, for this equality to hold, we get:
\begin{align} \label{greens_func_relations}
\tilde{\bar{\bar{G}}}^T(\mathbf{r}^\prime, \mathbf{r}) = \bar{\bar{G}}(\mathbf{r}, \mathbf{r}^\prime)
\end{align}
which applies to both nonreciprocal and reciprocal systems in thermodynamic quasi-equilibrium. This relation can also be derived using Onsager's reciprocal relations, as done in Ref. \cite{viktar_tutorial_EM_nonreciprocity_2020}.

\subsection{Fluctuational electrodynamics formalism}
In this section, we describe our theoretical model of scintillation within the framework of fluctuational electrodynamics.
Since thermalization occurs on a much faster time scale than the radiative stage of scintillation, the process can be described as being in quasi-equilibrium. As such, the fluctuation-dissipation theorem can be applied, as proposed in Ref. \cite{light_emission_nonequi_bodies_greffet_PRX_2018}. 




The current-current correlation function describing the scintillating current sources can be expressed as follows \cite{SM,quantum_noise_RevModPhys, light_emission_nonequi_bodies_greffet_PRX_2018}: 
\begin{align} 
    \langle &J_m(\mathbf{r}^{\prime}, \omega ) J_n^* (\mathbf{r}^{\prime \prime}, \omega^\prime ) \rangle \nonumber
\\ &= 2 \pi  \epsilon_0 \omega [\hbar \omega S(\omega)] \text{Im}[\bar{\bar{\epsilon}}_{mn} ]\delta(\omega-\omega^\prime) \delta(\mathbf{r}^{\prime} - \mathbf{r}^{\prime \prime}) \label{FDT}
\end{align}
where $m,n \in \{x,y,z\}$ are the components of the current density operator. 
 The imaginary part of the permittivity tensor $\text{Im}[\bar{\bar{\epsilon}}_{mn}]$ is proportional to the rate of absorption subtracted by the rate of stimulated emission and thus depends on the electronic states of the scintillating material \cite{light_emission_nonequi_bodies_greffet_PRX_2018, wurfel1982chemical}. Note that $\text{Im}[\bar{\bar{\epsilon}}]$ is not extracted from a state of true thermal equilibrium.
%
Here, $S(\omega)$ is the average occupation of an optical mode at frequency $\omega$ in the non-equilibrium steady state and can be approximated by fitting the emission spectra of the scintillator to a Gaussian, for example \cite{SM, yb:yag_emission_spectra_data_2006}. It is analogous to the Bose-Einstein distribution $\Theta(\omega) = 1/[e^{\hbar \omega/k_B T} - 1]$ in the case of thermal radiation.
Using the current-current correlation in Eq. \ref{FDT}, we can then compute the direct emission from the scintillating sources within the framework of fluctuational electrodynamics \cite{rytov1989_vol3}. 

Now using our derived Green's function relations in Eq. \ref{greens_func_relations} and the expression for $\langle J_m(\mathbf{r}^{\prime}, \omega ) J_n^* (\mathbf{r}^{\prime \prime}, \omega^\prime ) \rangle$ in Eq. \ref{FDT}, we can relate the fields emitted by our nonreciprocal structure to the fields in the adjoint structure. Beginning with the ensemble-averaged Poynting flux at $\mathbf{r}$, we have:
\begin{align}
    \langle \mathbf{S}(\mathbf{r},t)  \rangle &= \langle \mathbf{E}(\mathbf{r}, t) \times \mathbf{H}(\mathbf{r}, t) \rangle
\end{align}
Assuming $\langle\mathbf{S}(\mathbf{r},t)  \rangle$ is independent of time, its spectral density $\langle \mathbf{S}(\mathbf{r},\omega)  \rangle $ can be expressed as \cite{optical_coherence_quantum_optics_mandel_wolf}:
\begin{align}
    \langle \mathbf{S}(\mathbf{r},\omega)  \rangle &= \frac{1}{\pi^2} \int_{0}^{\infty} d\omega \bigg[ \frac{1}{2} \text{Re}[\langle \mathbf{E}(\mathbf{r},\omega) \times \mathbf{H}^*(\mathbf{r},\omega) \rangle] \bigg] \label{final_poynting_expression_main_text}
\end{align}
Since $\mathbf{H}$ can be expressed in terms of the $\mathbf{E}$ field through Maxwell's equations (i.e. $\mathbf{H} = (\nabla \times \mathbf{E})/i\mu \omega$ in free space), the Poynting flux given in Eq. \ref{final_poynting_expression_main_text} can then be computed with knowledge of the correlation function of the Fourier transform of the electric fields:
\begin{align}
    \langle &E_k(\mathbf{r}, \omega) E_l^*(\mathbf{r}, \omega) \rangle \nonumber \\ &= 2\pi \frac{\omega^3}{\epsilon_0 c^4} [ \hbar \omega S(\omega)] \int_{V^\prime} d\mathbf{r}^{\prime} \frac{\tilde{E}_{m}(\mathbf{r}^{\prime}, \omega)}{\omega^2 \mu_0 } \frac{\tilde{E}_{n}^*(\mathbf{r}^{\prime}, \omega)}{\omega^2 \mu_0 }   \frac{\epsilon_{mn} - \epsilon_{nm}^*}{2i} \label{intens_spectrum_with_E_fields} 
\end{align} 
where $\tilde{E}_{m}(\mathbf{r}^{\prime}, \omega)$ and $\tilde{E}_{n}(\mathbf{r}^{\prime}, \omega)$ are the $m$, $n$ component of the field induced in the adjoint structure by the $k$, $l$ component of a current density in the far-field, respectively.
The derivations can be found in Supplementary Materials (SM) \cite{SM}.

Thus, we expect that at a given frequency, the emission angles $\pm \theta$ that exhibit large differences in scintillation power (i.e. high degree of nonreciprocity) will also exhibit large differences in electric field concentrations in the adjoint structure when excited by plane waves incident from angles $+\theta$ and $-\theta$.


\subsection{Nonreciprocity for directional scintillation}

Scintillators are characterized by the absolute light yield (photons generated from an absorbed amount of ionizing radiation) and light output (light collected by the photodetector) \cite{Lecoq2020, review_of_scintillator_eng_2024}. In certain radiation detection schemes, photodetectors are located off-normal from the radiation source (this is used, for instance, to shield detection electronics from ionization radiation) \cite{scint_counting_birks_textbook_1964}. The energy resolution of such systems is improved by maximizing the light collection efficiency: the ratio of light output to absolute light yield \cite{Lecoq2020}.
Thus, it is often desirable to enhance the scintillation emission signal along an off-normal direction, toward the photodetector \cite{zhu_enhancement_of_directional_scintillation_2017}. Previous designs for such directional control of scintillation include placing photonic crystal structures on top of the scintillator and microlens arrays, which have all been reciprocal \cite{zhu_enhancement_of_directional_scintillation_2017, directional_scint_microlens_array_2021}. 

\begin{figure}
\includegraphics[width=0.45\textwidth]{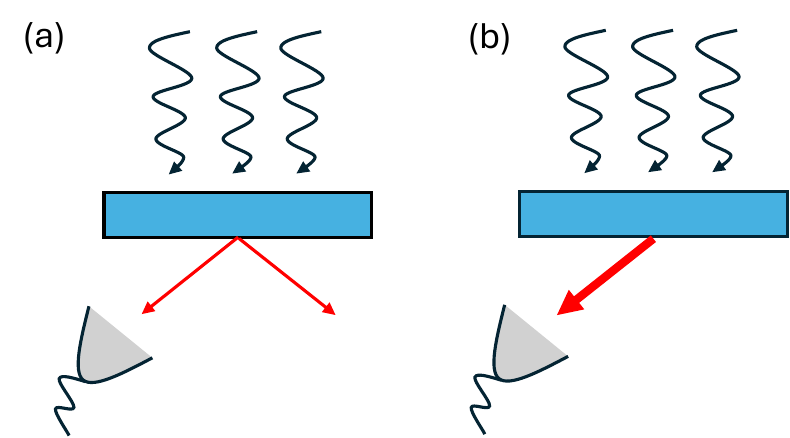}
\caption{\label{direc_emission} Directional emission in (a) reciprocal and (b) nonreciprocal system. Photodetector is located along an off-normal direction. Red arrows indicate scintillation emission and thickness corresponds to emission intensity.}
\end{figure}

As discussed in Section IIA, $\varepsilon (\theta) = \varepsilon (-\theta)$ in a reciprocal, two-port system   (Fig. \ref{direc_emission}a). In such a system, a strong emission along the $+\theta$ direction necessarily implies a strong emission along the $-\theta$ direction. In other words, with reciprocity, it is not possible to make such a system emit only to one angle. In contrast, by breaking reciprocity, it is possible to achieve a strong angular asymmetry, i.e. $\varepsilon (\theta) = 0, \varepsilon(-\theta) \neq 0$ (Fig. \ref{direc_emission}b).
%
%
Therefore, one can create a scintillator that emits along a single direction. In such a case, it is in principle possible to achieve perfect light collection efficiency for scintillation emission at an off-normal direction. Moreover, compared to the reciprocal case for a given incident power, the peak emission intensity is enhanced in the nonreciprocal case since the number of emission channels can be reduced. 

\section{Numerical results}

%
Motivated by the theoretical discussion in the previous section, in this section, we design a magneto-optical photonic crystal structure that exhibits nonreciprocal scintillation and provide numerical simulations of its emission characteristics.
Magneto-optical grating structures have been previously designed for nonreciprocal thermal radiation applications \cite{linxiao_near-complete_violation_PRB_2014, bo_zhao_near-complete_violation_kirchhoff_opt_letters_2019}. 
Here, motivated by recent works on scintillator designs using multiple planar layers, we consider a magneto-optical photonic crystal consisting of alternating layers of scintillating material and magneto-optic material.

To achieve the nonreciprocal effects of angular asymmetry in such a planar structure, the structure must break both space-inversion symmetry and time-reversal symmetry \cite{nonrecip_mag_phc_PRE_2001}. Thus, a one-dimensional photonic crystal with only two layers per unit cell cannot exhibit angular asymmetry due to space-inversion symmetry \cite{nonrecip_mag_phc_PRE_2001, one-way_total_reflec_Fan_2007}. In light of this, we design the simplest unit cell consisting of 3 layers to break inversion symmetry. Time-reversal symmetry is broken by the magneto-optic material with an applied external magnetic field. 

\begin{figure*}
\includegraphics[width=0.8\textwidth]{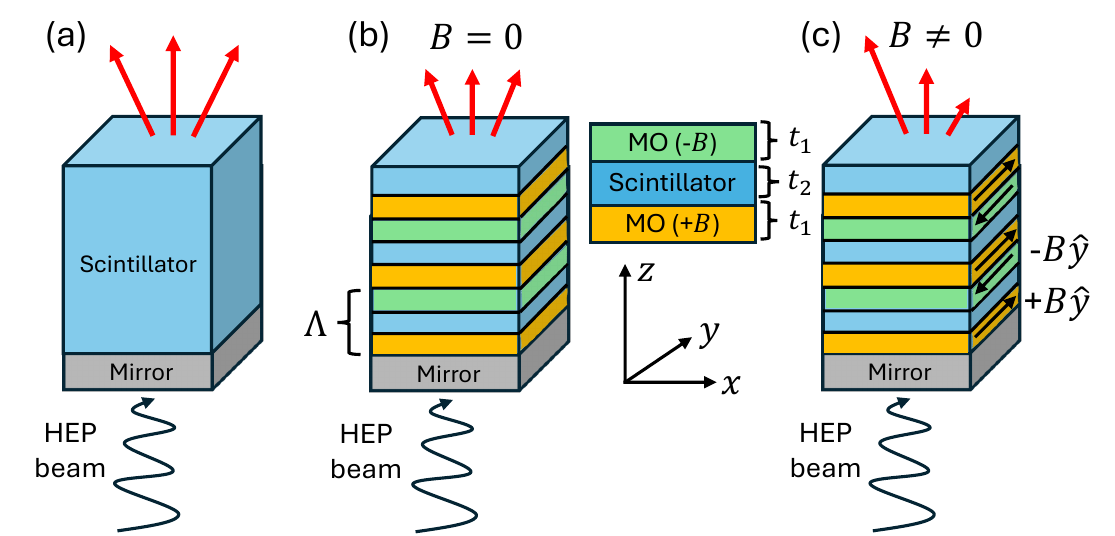}
\caption{\label{structure_fig} Schematics of structures for comparison: (a) bulk scintillator (b) magneto-optical photonic crystal with no external $\mathbf{B}$ field and (c) with applied external $\mathbf{B}$ field. Yellow layers indicate magneto-optic material with $\mathbf{B} = +B\hat{y}$ and green layers indicate magneto-optic (MO) material with  $\mathbf{B} = -B\hat{y}$. A unit cell with period $\Lambda$ consists of a scintillating layer between two layers of magneto-optic material with $\mathbf{B}$ fields in opposite directions, as shown in inset of (b). The layer thicknesses are $t_1=82$nm and $t_2=50$nm. Scintillation is excited by a high-energy particle (HEP) beam.}
\end{figure*}

\begin{figure*}
\includegraphics[width=1\textwidth]{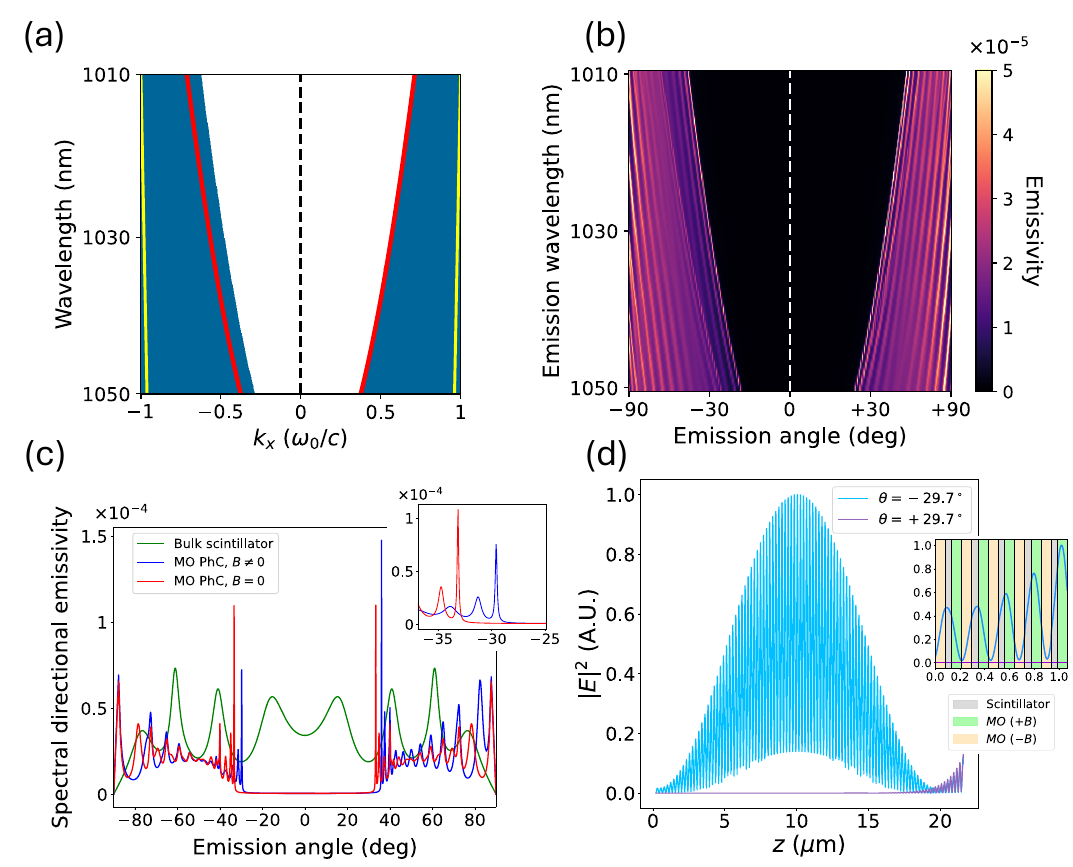}
\caption{\label{nonrecip_radiation} (a) Projected band structure of infinite magneto-optic photonic crystal for TM polarization. Red lines indicate the band edge for $+k_x$ and its reflection with respect to 0 to demonstrate the asymmetry of the band structure. Yellow lines indicate the light line. $\omega_0$ is the highest frequency in the emission bandwidth. (b) Emissivity plot for unpolarized emission from multilayer structure; Dashed line at $\theta = 0$ to show asymmetry
(c) Spectral directional emissivity of unpolarized light at peak emission wavelength 1030nm for the 3 setups shown in Fig. \ref{structure_fig}.
(d)
Spatial distribution of $|E|^2$ fields in adjoint structure under TM plane wave excitation from $z = \infty$ and from incident angles $\theta= \pm 29.7^\circ$ at the wavelength $\lambda=1030$ nm. Values are normalized by the maximum value of the $\theta=-29.7^\circ$ plot. The value $|\theta|=29.7^\circ$ is where maximum emission occurs for $\theta<0$ values and resides in the bandgap for $\theta>0$ values. Inset shows $|E|^2$ fields in bottom 5 unit cells of the structure.
Field plots were computed using the frequency-domain solver of COMSOL Multiphysics\textsuperscript{\textregistered}. 
}
\end{figure*}

In our design, we consider the Voigt configuration \cite{modern_magnetooptics_book_zvezdin_1997}, where the external magnetic field $\mathbf{B}$ is applied perpendicular ($\pm \hat{y}$) to the direction of propagation (Fig. \ref{structure_fig}c). In this configuration, the TM and TE polarizations are decoupled, and the TM polarization experiences the nonreciprocal effects \cite{one-way_total_reflec_Fan_2007}.
With the applied external $\mathbf{B}$ field, the relative permittivity tensor of the magneto-optic material takes the form: 
\begin{equation}
\bar{\bar{\epsilon}} = \begin{pmatrix} \epsilon_d & 0 & ig \\
0 & \epsilon_d & 0 \\
-ig & 0 & \epsilon_d
\end{pmatrix}
\end{equation}
For $\mathbf{B} = +B\hat{y}$ and $\mathbf{B} = -B\hat{y}$, $g > 0$ and $g < 0$, respectively. In our design, we use $\epsilon_d = 6.25$ and $g= \pm 0.3$ \cite{theory_paper_using_g=0.3_BIG}, which approximates bismuth iron garnet (BIG) at near-infrared wavelengths. For the scintillating material, we use Yb:YAG, which has relative permittivity Re$(\epsilon_s) = 3.295$ at emission wavelengths $\lambda = 1010-1050$nm \cite{refrac_index_YAG_undoped_1998, yb:yag_emission_spectra_data_2006}. 
The 3-layer unit cell consists of a $50$nm layer of scintillator between two $82$nm layers of magneto-optic material with the external magnetic field applied in opposite directions ($\pm \hat{y}$) to achieve  maximal nonreciprocity \cite{magneto-optic_defects_in_phc_2005}. Such a design can be implemented using ferrimagnetic materials with magnetic domains that have opposite magnetization directions \cite{review_magnetic_phc_2003}. 
 The projected band structure of the TM polarization for the infinite magneto-optic photonic crystal is shown in Fig. \ref{nonrecip_radiation}a. We see that there is clear asymmetry with respect to $k_x = 0$. At a given frequency, the structure may support photonic modes at $+k_x$ but not at $-k_x$. Such an asymmetry in the band structure should translate into angular asymmetry in the emission properties of the structure. 


In general, the spatial distribution of the scintillating current sources will depend on the energy loss density of the incident high-energy particle beam. 
In our simulations, 
we assume a uniform distribution of emitters in the $z$-direction, which is valid when the incident particle beam can penetrate through the structure entirely. We model the emission from a finite-sized device with 100 periods on top of a reflecting aluminum mirror, as depicted schematically in Fig. \ref{structure_fig}c. Since our photonic crystal structure has a total thickness of $21.4$$\mu$m, the assumption of uniform emitter distribution is reasonable for incident photon energies in the range $10^{-1}-10^4$ MeV, as detailed in SM \cite{SM}. The mirror is used to achieve a two-port system as described in Section II and is typically used in scintillator detectors to couple more scintillation light to the detector. 


Using rigorous coupled wave analysis (RCWA) combined with the fluctuational electrodynamics formalism, we compute the spectral directional emissivity of unpolarized light in the Voigt configuration in Fig. \ref{nonrecip_radiation}b \cite{kaifeng_mesh_2018, cuevas_s-matrix_magneto-optic_2012, culshaw_1999}. The plotted frequencies cover the emission bandwidth of the scintillator Yb:YAG. 
We use the value Im$(\epsilon_s) = 1$e$-6$, which depends on the flux of the incident X-ray. Our choice of this value here is consistent with experiment \cite{framework_scintillation_science_2022, SM}. 
As noted in Section IIC, this value of Im$(\epsilon_s)$ differs from that of the permittivity of the material in thermal equilibrium, since it corresponds to the scintillator in a non-equilibrium state (under X-ray excitation).
From the plot, we observe the angular asymmetry in the emissivity, confirming the nonreciprocal behavior of the system. The emissivity plot agrees well with the asymmetric band structure of the infinite magneto-optic photonic crystal in Fig. \ref{nonrecip_radiation}a. 
We note that the emissivity plot of Fig. \ref{nonrecip_radiation}b is plotted with respect to the emission angle. This corresponds to the band structure in Fig. \ref{nonrecip_radiation}a that lies above the light line $\omega = c k_x$, as indicated by the yellow line in Fig. \ref{nonrecip_radiation}a.





In Fig. \ref{nonrecip_radiation}c, we plot the spectral directional emissivity at the wavelength $\lambda = 1030$nm, the peak emission wavelength of the scintillator Yb:YAG. We compare the emissivities of the three structures shown in Fig. \ref{structure_fig}: the nonreciprocal multilayer structure with and without an applied external magnetic field, as well as the bulk scintillator with thickness $5$$\mu$m, which is the total thickness of scintillating material in the multilayer structure. We note that the Fabry-Perot oscillations are visible in the directional emissivity of the bulk scintillator. 
In the cases of the bulk scintillator and the multilayer structure without an applied magnetic field, the spectral directional emissivity is symmetric with respect to normal incidence (i.e. $\varepsilon(+\theta)=\varepsilon(-\theta)$), as expected due to reciprocity. With an applied magnetic field, we find that the emissivity curve becomes asymmetric. 
As can be seen in Fig. \ref{nonrecip_radiation}c, for the multilayer structure, with or without magnetic field, there is an angular range $[-\theta^-,\theta^+]$ around the normal direction of $\theta=0$, for which the emissivity is near zero. (In this notation both $\theta^-$ and $\theta^+$ are positive). This angular range is a result of the photonic band gap of the system \cite{Winn_98}.
In Section IIC, we theoretically relate the emitted scintillation power in the directions $\pm \theta$ to the fields induced in the adjoint structure by plane waves incident from $\pm \theta$.
In Fig. \ref{nonrecip_radiation}d, we show the magnitude of the absorbed electric field energy as a function of depth in the adjoint structure. We excite the structure with TM plane waves from $z = \infty$ with incident angles $\theta = \pm 29.7^\circ$, which lie within the angular range $\theta^- < |\theta| < \theta^+$ where strong angular asymmetry occurs. We see that the incident fields from $+29.7^\circ$ decay rapidly within the structure, whereas the fields from $-29.7^\circ$ are present throughout the structure. This high asymmetry in the field concentrations is directly related to the strong nonreciprocal angular asymmetry in the emission at this angle, as predicted in Eq. \ref{intens_spectrum_with_E_fields} of Section IIC. 
%
%

%
%

\begin{figure*}
\includegraphics[width=0.9\textwidth]{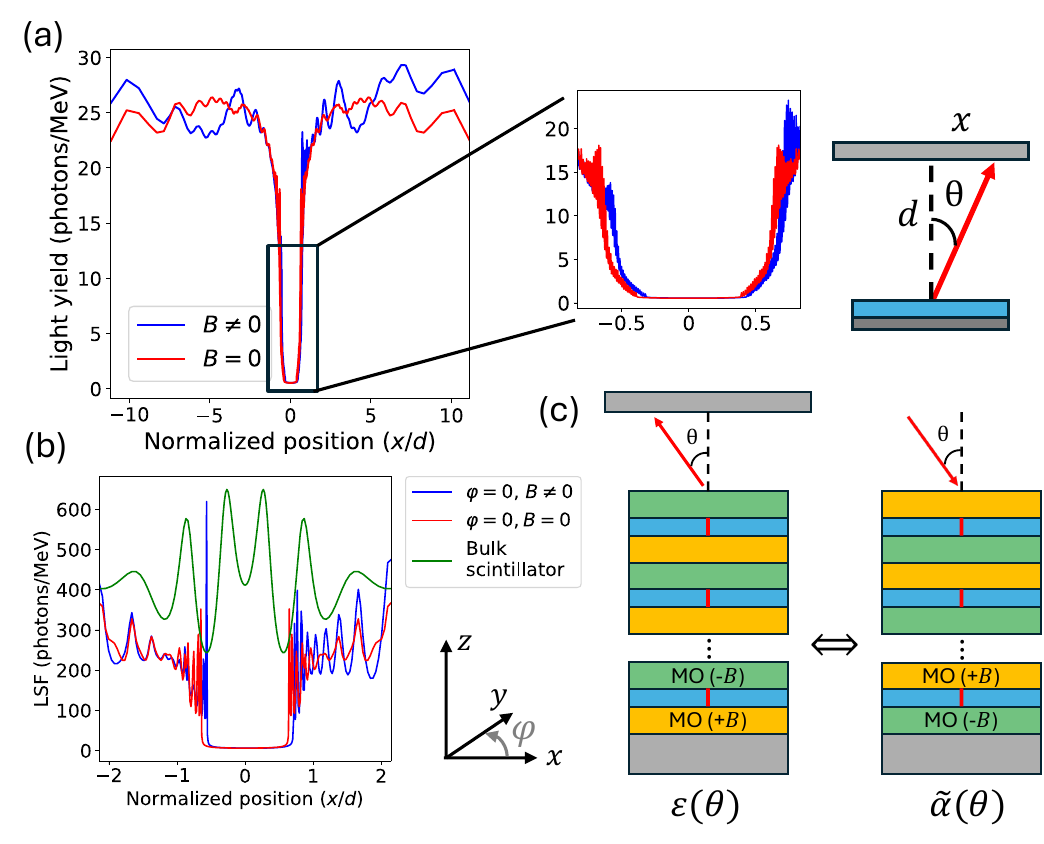}
\caption{\label{scint_fig} (a) Scintillation light yield of magneto-optical photonic crystal structure as function of emission angle for $\mathbf{B} \neq 0$ and $\mathbf{B}=0 $ cases. Inset shows zoom-in on minimum emission angles for the two cases resulting from nonreciprocity. Far right inset shows relation between emission angle $\theta$ to position $x$ on detector plane located a distance $d$ away from structure. (b) Line spread function (LSF) plot along $\varphi=0$ direction for multilayer structure (with and without applied $\mathbf{B}$ field) and bulk scintillator. The integrated LSF of the bulk is normalized to match the experimental light yield value. (c) Schematic of equivalence between an emitting sheet of sources at the plane $x_0=0$ and the absorption along the same plane in the adjoint structure, based on adjoint Kirchhoff's law. }
\end{figure*}

We now examine the emission of the structure integrated over the emission bandwidth of the scintillator Yb:YAG. In Fig. \ref{scint_fig}a, we plot the expected light yield as a function of position on the detector for the case when $\mathbf{B} = 0$ and $\mathbf{B} \neq 0$. To compute this plot, we first obtain the spectral flux density $\Phi(\theta)$ of emitted light summed over both polarizations and integrated over the emission bandwidth of the scintillator material Yb:YAG, given by the expression:
\begin{align}
    \Phi(\theta) = \alpha \sum_\omega \varepsilon(\omega, \theta) \hbar \omega S(\omega)
\end{align}
where $S(\omega)$ is the occupation probability of a mode at frequency $\omega$ in the scintillator, $\varepsilon(\omega, \theta)$ are the computed spectral emissivities from Fig. \ref{nonrecip_radiation}b, and $\alpha$ is a normalization factor used to match the experimental value of light yield for Yb:YAG under electron beam excitation ($79 \times 10^3$ photons/MeV \cite{infrared_scint_Yb_YAG_e_beam_light_yield}).
$S(\omega)$ is obtained by fitting to the spectral emission cross section of Yb:YAG \cite{SM}.
To convert the emission angles to positions at the detector plane, we assume the detector plane is located at a distance of $d$ from the structure (right inset of Fig. \ref{scint_fig}).
We see that computed light yield becomes asymmetric around $x = 0$, in the presence of the magnetic field. Thus, the nonreciprocal angular asymmetry should persist in the detected signal, even after spectral integration.

In imaging applications, it is important to consider point and line spread functions since these functions characterize the spatial resolutions of the system \cite{medical_imaging_text}. As an illustration, for our system, consider a ray of high energy particles incident perpendicular to the scintillator at an in-plane location $(x_0, y_0)$, producing light emission at various depth positions $z$ in the scintillator. All these emissions are then detected. The light distribution at the detector plane as parameterized by the coordinates $(x,y)$ then defines the point spread function $h(x-x_0,y-y_0)$. Analogously, a line spread function describes the emission as produced by a sheet of incident high energy particles.  The line spread function can be related to the point spread function. For example, if the sheet is located at an in-plane location $x_0$,  the line spread function is then \cite{computing_PSF_from_LSF_for_arbitrary_system_Marchand_1965, PSF_LSF_MTF_radiology_1969}:
\begin{align}
    \text{LSF}(x-x_0) = \int_{-\infty}^{\infty} h(x-x_0,y-y_0) dy_0 \label{lsf_defn}
\end{align}
Note that on the right-hand side, the result of the integration is independent of $y$ by translational symmetry. 


In Fig. \ref{scint_fig}b, we plot the LSF of Eqn. \ref{lsf_defn} in terms of the light yield at the operating wavelength $\lambda = 1030$nm, the emission peak wavelength of Yb:YAG. The LSF is computed by first obtaining the absorbed power along the plane of $x_0 = 0$ in the adjoint structure excited by an incident plane wave, which is equivalent to the integrated emission from all sources at the plane $x_0 = 0$ in the original structure by the adjoint Kirchhoff's law, as shown schematically in Fig. \ref{scint_fig}c \cite{adjoint_kirchhoff_law_cheng_PRX_2022, david_miller_universal_radiation_laws_PNAS_2017}. Numerical confirmation of the adjoint Kirchhoff's law can be found in SM \cite{adjoint_kirchhoff_law_cheng_PRX_2022, SM}. The integrated LSF of the bulk scintillator was normalized to match the experimental light yield value for Yb:YAG. This normalization procedure was also applied to the multilayer LSF plots. 
The resulting LSF is directly related the angular spectrum of the emissivity as shown in Fig. \ref{nonrecip_radiation}c, up to a normalization factor, provided that we convert $x/d$ to the emission angle. In general, the emissivity spectrum takes into account the sources in the entire structure, whereas the LSF considers only the sources on a sheet. In the present case, however, they are proportional to each other due to translational symmetry. The nonreciprocity manifests also in the LSF in terms of the asymmetry between $x$ and $-x$. 

%

\section{Discussion}
We conclude this work with a few remarks. First of all, as mentioned in the text, in our structure the directional scintillation emission is a direct consequence of reciprocity breaking, since our system, being planar, is a two-port system. We note that in general, for systems with more than two ports, the directional emission can be achieved with reciprocal systems \cite{coherent_emission_light_by_thermal_sources_greffet_2002, highly_direc_radiation_tungsten_2005, thermal_metasurf_alu_PRX_2021}, for example, with the use of asymmetric grating structures having periodicity larger than the wavelength \cite{asymm_directional_control_thermal_2023}. Here, we focus on a planar system since most scintillator systems are planar \cite{inorganic_scint_text_2017}, and also because such a planar system provides a direct signature in emission of nonreciprocity. Also, in our system, the emission pattern can be tuned by adjusting the strength of the magnetic field, which may provide a mechanism for dynamic control of scintillation emission. In addition, in our system, we use a substantial number of layers so that the nonreciprocal contrast between different directions is large. Nonreciprocal emission can be seen with far fewer layers, as detailed in Section III of the SM \cite{SM}. Alternatively, one may consider grating structures with sub-wavelength periodicity, which can also be described as a two-port system \cite{linxiao_near-complete_violation_PRB_2014}, as a way to achieve nonreciprocal scintillation emission. 

In this work, we focus on the use of magneto-optical effects to break reciprocity. For future work, it would be of interest to explore other mechanisms of reciprocity breaking, such as the use of time-modulation \cite{zongfu_optical_isolation_interband_2009} or nonlinearity \cite{nonlinear_nonrecip_alu_2021} in the context of the control of scintillation. One may further extend our theoretical framework to include the use of the chemical potential of light generated by scintillation to derive the mode occupation function \cite{wurfel1982chemical}. In addition, the effects of a non-uniform spatial distribution of emitting scintillation sources on nonreciprocal emission characteristics is an interesting topic of future study.

\section{Summary}
In summary, we have proposed the design of a nonreciprocal scintillator using magneto-optical photonic crystals.  
%
%
Our work demonstrates the potential of controlling non-equilibrium radiation through breaking reciprocity. Such nonreciprocity can be harnessed to enhance directional scintillation emission by reducing the number of angular emission channels, thus finding potential applications in various radiation detection schemes.



\begin{acknowledgments}
We wish to acknowledge Dr. Cheng Guo, Dr. Yubin Park, and Prof. Linxiao Zhu for helpful discussion. C.~R.-C. is supported by a Stanford Science Fellowship. S. P. gratefully acknowledges support from the NSF GRFP under Grant No. 2141064. O. Y. L. is supported by a Stanford Graduate Fellowship. This publication was also supported in part by the DARPA Agreement No. HO0011249049.
\end{acknowledgments}

\bibliography{bibliography}

\ifarXiv
    

\section*{Supplementary material}

# Supplementary Material for: "Nonreciprocal scintillation using one-dimensional magneto-optical photonic crystals"

Olivia Y. Long,[1, *] Simo Pajovic,[2] Charles Roques-Carmes,[3] Yoichiro Tsurimaki,[3]
Nicholas Rivera,[4] Marin Soljačić,[5] Svetlana V. Boriskina,[2] and Shanhui Fan[1, 3, †]

[1]*Department of Applied Physics, Stanford University, Stanford, California 94305, USA*
[2]*Department of Mechanical Engineering, Massachusetts Institute of Technology, Cambridge, Massachusetts 02139, USA*
[3]*E. L. Ginzton Laboratory and Department of Electrical Engineering,*
*Stanford University, Stanford, California 94305, USA*
[4]*Department of Physics, Harvard University, Cambridge, Massachusetts 02138, USA*
[5]*Department of Physics, Massachusetts Institute of Technology, Cambridge, Massachusetts 02139, USA*
(Dated: September 25, 2024)

## CONTENTS

## I. DERIVATION OF SPECTRAL FIELD CORRELATION FUNCTION

Here, we derive the relation between the emitted fields from our nonreciprocal structure and the absorbed fields in the adjoint structure. We start with the expression for the ensemble-averaged Poynting flux at $\mathbf{r}$:

$$\langle \mathbf{S}(\mathbf{r}, t) \rangle = \langle \mathbf{E}(\mathbf{r}, t) \times \mathbf{H}(\mathbf{r}, t) \rangle \tag{1}$$

where $\mathbf{E}(\mathbf{r}, t)$ and $\mathbf{H}(\mathbf{r}, t)$ are the real electric and magnetic fields. Using the Fourier convention: $A(t) = \frac{1}{2\pi} \int e^{-i\omega t} A(\omega) d\omega$, we have:

$$\langle \mathbf{E}(\mathbf{r}, t) \times \mathbf{H}(\mathbf{r}, t) \rangle = \frac{1}{(2\pi)^2} \int_{-\infty}^{\infty} d\omega e^{-i\omega t} \int_{-\infty}^{\infty} d\omega' e^{+i\omega' t} \langle \mathbf{E}(\mathbf{r}, \omega) \times \mathbf{H}^*(\mathbf{r}, \omega') \rangle \tag{2}$$

Assuming $\langle \mathbf{E}(\mathbf{r}, t) \times \mathbf{H}(\mathbf{r}, t) \rangle$ is independent of time, we have:

$$\langle \mathbf{E}(\mathbf{r}, \omega) \times \mathbf{H}^*(\mathbf{r}, \omega') \rangle = F(\omega) \delta(\omega - \omega') \tag{3}$$

Below, in a slight but useful abuse of notation, we denote $F(\omega)$ as $\langle \mathbf{E}(\mathbf{r}, \omega) \times \mathbf{H}^*(\mathbf{r}, \omega) \rangle$. Then, we have:

$$\langle \mathbf{S}(\mathbf{r}, \omega) \rangle = \frac{1}{(2\pi)^2} \int_{-\infty}^{\infty} d\omega \langle \mathbf{E}(\mathbf{r}, \omega) \times \mathbf{H}^*(\mathbf{r}, \omega) \rangle \tag{4}$$

* olong@stanford.edu
† shanhui@stanford.edu

Since the fields are real, we know that $\mathbf{E}(\mathbf{r}, \omega) = \mathbf{E}^*(\mathbf{r}, -\omega)$ and $\mathbf{H}(\mathbf{r}, \omega) = \mathbf{H}^*(\mathbf{r}, -\omega)$. Thus, Eq. 4 can be expressed as:

$$\langle \mathbf{S}(\mathbf{r}, \omega) \rangle = \frac{1}{(2\pi)^2} \int_0^\infty d\omega \langle \mathbf{E}(\mathbf{r}, \omega) \times \mathbf{H}^*(\mathbf{r}, \omega) \rangle + \int_{-\infty}^0 d\omega \langle \mathbf{E}(\mathbf{r}, \omega) \times \mathbf{H}^*(\mathbf{r}, \omega) \rangle \tag{5}$$

$$= \frac{1}{(2\pi)^2} \int_0^\infty d\omega \left[ \langle \mathbf{E}(\mathbf{r}, \omega) \times \mathbf{H}^*(\mathbf{r}, \omega) \rangle + \langle \mathbf{E}(\mathbf{r}, -\omega) \times \mathbf{H}^*(\mathbf{r}, -\omega) \rangle \right] \tag{6}$$

$$= \frac{1}{(2\pi)^2} \int_0^\infty d\omega \left[ \langle \mathbf{E}(\mathbf{r}, \omega) \times \mathbf{H}^*(\mathbf{r}, \omega) \rangle + \langle \mathbf{E}^*(\mathbf{r}, \omega) \times \mathbf{H}(\mathbf{r}, \omega) \rangle \right] \tag{7}$$

$$= \frac{1}{(2\pi)^2} \int_0^\infty d\omega \left[ 2\mathrm{Re}[\langle \mathbf{E}(\mathbf{r}, \omega) \times \mathbf{H}^*(\mathbf{r}, \omega) \rangle] \right] \tag{8}$$

$$= \frac{1}{\pi^2} \int_0^\infty d\omega \left[ \frac{1}{2} \mathrm{Re}[\langle \mathbf{E}(\mathbf{r}, \omega) \times \mathbf{H}^*(\mathbf{r}, \omega) \rangle] \right] \tag{9}$$

Since $\mathbf{H}$ can be expressed in terms of the $\mathbf{E}$ field through Maxwell's equations (i.e. $\mathbf{H} = (\nabla \times \mathbf{E})/i\mu\omega$ in free space), the Poynting flux can be computed from Eq. 9 with knowledge of the correlation function of the electric field $\langle E_k(\mathbf{r}, \omega) E_l^*(\mathbf{r}, \omega) \rangle$. Thus, we now focus on the quantity $\langle E_k(\mathbf{r}, \omega) E_l^*(\mathbf{r}, \omega) \rangle$ in our nonreciprocal scintillating structure, which we can show to be related to the fields in the adjoint structure.

We can express components of the electric field in terms of the $m$ component of the current density $J_m(\mathbf{r}', \omega)$ at $\mathbf{r}'$: $E_k(\mathbf{r}, \omega) = i\mu_0\omega \int_{V'} d\mathbf{r}' G_{km}(\mathbf{r}, \mathbf{r}', \omega) J_m(\mathbf{r}', \omega)$, where $G_{km}(\mathbf{r}, \mathbf{r}', \omega)$ is the $k$ component of the Green's function at $\mathbf{r}$ induced by the $m$ component of the current density.

Since thermalization occurs on a much faster time scale than spontaneous emission in the scintillation process, as noted in the main text, we treat the state as a quasi-equilibrium state and apply the fluctuation-dissipation theorem to obtain the expression for $\langle J(\mathbf{r}', \omega) \cdot J^*(\mathbf{r}'', \omega) \rangle$ [1], which is determined by the energy levels of the scintillating electrons and the occupation of those levels:

$$\langle J_m(\mathbf{r}', \omega) J_n^*(\mathbf{r}'', \omega') \rangle = 2\pi\epsilon_0\omega[\hbar\omega S(\omega)] \frac{\epsilon_{mn} - \epsilon_{nm}^*}{2i} \delta(\omega - \omega') \delta(\mathbf{r}' - \mathbf{r}'') \tag{10}$$

where we have assumed that the permittivity is local. The term $[\hbar\omega S(\omega)]$ is the average energy of a photonic mode with frequency $\omega$ in the quasi-equilibrium state. $S(\omega)$ is the occupation probability of a mode with frequency $\omega$ (see Section VI). In the case of thermal radiation, the term $[\hbar\omega S(\omega)] = \Theta(\omega, T) = \hbar\omega/[\exp[\hbar\omega/k_B T] - 1]$.

As before, we denote:

$$\langle J_m(\mathbf{r}', \omega) J_n^*(\mathbf{r}'', \omega') \rangle = \langle J_m(\mathbf{r}', \omega) J_n^*(\mathbf{r}'', \omega) \rangle \delta(\omega - \omega')$$

Then, we have:

$$\langle E_k(\mathbf{r}, \omega) E_l^*(\mathbf{r}, \omega) \rangle = (\mu_0\omega)^2 \int_{V'} d\mathbf{r}' \int_{V''} d\mathbf{r}'' G_{km}(\mathbf{r}, \mathbf{r}', \omega) G_{ln}^*(\mathbf{r}, \mathbf{r}'', \omega) \langle J_m(\mathbf{r}', \omega) J_n^*(\mathbf{r}'', \omega) \rangle \tag{11}$$

Plugging Eq. 10 into Eq. 11, we obtain:

$$\langle E_k(\mathbf{r}, \omega) E_l^*(\mathbf{r}, \omega) \rangle = \mu_0^2\omega^2 \int_{V'} d\mathbf{r}' \int_{V''} d\mathbf{r}'' G_{km}(\mathbf{r}, \mathbf{r}', \omega) G_{ln}^*(\mathbf{r}, \mathbf{r}'', \omega) 2\pi\epsilon_0\omega[\hbar\omega S(\omega)] \frac{\epsilon_{mn} - \epsilon_{nm}^*}{2i} \delta(\mathbf{r}' - \mathbf{r}'') \tag{12}$$

$$= 2\pi\epsilon_0\mu_0^2\omega^3[\hbar\omega S(\omega)] \int_{V'} d\mathbf{r}' \int_{V''} d\mathbf{r}'' G_{km}(\mathbf{r}, \mathbf{r}', \omega) G_{ln}^*(\mathbf{r}, \mathbf{r}'', \omega) \frac{\epsilon_{mn} - \epsilon_{nm}^*}{2i} \delta(\mathbf{r}' - \mathbf{r}'') \tag{13}$$

$$= 2\pi \frac{\omega^3}{\epsilon_0 c^4}[\hbar\omega S(\omega)] \int_{V'} d\mathbf{r}' G_{km}(\mathbf{r}, \mathbf{r}', \omega) G_{ln}^*(\mathbf{r}, \mathbf{r}', \omega) \frac{\epsilon_{mn} - \epsilon_{nm}^*}{2i} \tag{14}$$

(Note: $\mathbf{r}'$ is inside the scintillator structure and $\mathbf{r}$ is in the far-field.)

Using the generalized reciprocity theorem $\tilde{G}^T(\mathbf{r}', \mathbf{r}) = G(\mathbf{r}, \mathbf{r}')$ where $\tilde{G}$ is the Green's function of the complementary system with permittivity tensor $\overline{\overline{\epsilon}}^T$ [2], we can relate the fields radiated by sources in the scintillating structure (to the far-field) to the fields received by the adjoint structure (from radiating sources in the far-field).

Thus, Eq. 14 becomes:

$$\langle E_k(\mathbf{r}, \omega) E_l^*(\mathbf{r}, \omega) \rangle = 2\pi \frac{\omega^3}{\epsilon_0 c^4}[\hbar\omega S(\omega)] \int_{V'} d\mathbf{r}' \tilde{G}_{mk}(\mathbf{r}', \mathbf{r}, \omega) \tilde{G}_{nl}^*(\mathbf{r}', \mathbf{r}, \omega) \frac{\epsilon_{mn} - \epsilon_{nm}^*}{2i} \tag{15}$$

The Green's function at $\mathbf{r}'$ induced by a current density $J(\mathbf{r}, \omega) = -i\omega\delta(x)\delta(y)\delta(z)\hat{\mathbf{j}}$, where $\hat{\mathbf{j}}$ is the unit vector pointing along the $\mathbf{j}$ direction, is:

$$E_k(\mathbf{r}', \omega) = \omega^2\mu_0 \int_V d\mathbf{r} G_{kj}(\mathbf{r}', \mathbf{r}, \omega)\delta(x)\delta(y)\delta(z) \tag{16}$$

$$= \omega^2\mu_0 G_{kj}(\mathbf{r}', \mathbf{r}, \omega) \tag{17}$$

Plugging this into Eq. 15, we have:

$$\langle E_k(\mathbf{r}, \omega)E_l^*(\mathbf{r}, \omega)\rangle = 2\pi\frac{\omega^3}{\epsilon_0 c^4}[\hbar\omega S(\omega)] \int_{V'} d\mathbf{r}' \frac{\tilde{E}_m(\mathbf{r}', \omega)}{\omega^2\mu_0}\frac{\tilde{E}_n^*(\mathbf{r}', \omega)}{\omega^2\mu_0}\frac{\epsilon_{mn} - \epsilon_{nm}^*}{2i} \tag{18}$$

This expression can then be used to compute the Poynting flux of our structure, as discussed above. Thus, we have related the far-field emission by the nonreciprocal scintillating structure to the fields induced in the adjoint structure (from a current density in the far-field). To clarify, $\tilde{E}_m(\mathbf{r}', \omega)$ and $\tilde{E}_n(\mathbf{r}', \omega)$ are the $m$, $n$ component of the field induced in the adjoint structure by the $k$, $l$ component of a current density in the far-field, respectively.

### A. Note on extracting Im($\epsilon$) of the non-equilibrium state from experimental data

Im($\epsilon$) of the non-equilibrium state is proportional to the absorption of the scintillator, which was extracted by measuring the scintillation enhancement between the patterned and unpatterned portion of the photonic crystal in Ref. [3]. The scintillation enhancement is equal to the ratio of the absorption cross sections, which is equal to the ratio of the effective volumes $V_{eff}$ of the 2 structures. $V_{eff}$ depends on Im($\epsilon$) since Im($\epsilon$) is used to compute $V_{eff}$ in RCWA.

The value of Im($\epsilon$) extracted in this way is the value in non-equilibrium since it is obtained from the scintillation signal.

## II. CONFIRMATION OF ADJOINT KIRCHHOFF'S LAW

In Fig. 1a, we show the emissivity and the absorptivity for the TM polarization at $\lambda = 1030$nm for the nonreciprocal multilayer structure ($g = 0.3$). The emissivity was computed using RCWA within the framework of fluctuational electrodynamics [4]. The emission from only the scintillating layers was computed. The absorptivity was obtained separately by computing the absorbed electromagnetic power in the scintillating layers using COMSOL. We see that $e(\theta) \neq \alpha(\theta)$, as expected due to nonreciprocity. In Fig. 1b, we see that the emissivity of the adjoint structure matches the absorptivity of the original nonreciprocal multilayer structure, confirming the adjoint Kirchhoff's law.

## III. EMISSIVITY FROM FEWER PERIODS IN MULTILAYER STRUCTURE

In this section, we show that the emissivity contrast $\varepsilon(-\theta)/\varepsilon(+\theta)$ persists with fewer periods of our multilayer structure. In Fig. 2, we show the emissivity for unpolarized emission from 5, 10, 20, and 30 periods, which correspond to total thicknesses of 1.07 $\mu$m, 2.14 $\mu$m, 4.28 $\mu$m, and 6.42 $\mu$m. As shown, the emissivity contrast increases with the number of periods. Thus, to achieve higher emissivity contrast in our proof of principle design, we used a larger number of periods, but nonreciprocal emissivity contrast can be seen with a much smaller number of layers.

## IV. OPTIMIZATION OF LAYER THICKNESSES IN STRUCTURE

A structure was first obtained by optimizing for:

$$\min_{\mathbf{t}\in\mathbb{R}^3} \frac{1}{\max A(+\theta)/A(-\theta)} + \frac{1}{|\theta_{A^s,\max} - \theta_{A,\max}|} \tag{19}$$

$$\text{s.t.} \quad 0 < |\theta| < 30° \tag{20}$$

where $A^s$ is the symmetric absorptance values obtained by reflecting the $+\theta$ values to the $-\theta$ side.

In words, we maximize the nonreciprocity ratio $A(+\theta)/A(-\theta)$ and also maximize the difference in $\theta$ (on the $-\theta$ side) where the maximum values of $A^s$ vs. $A$ occur.

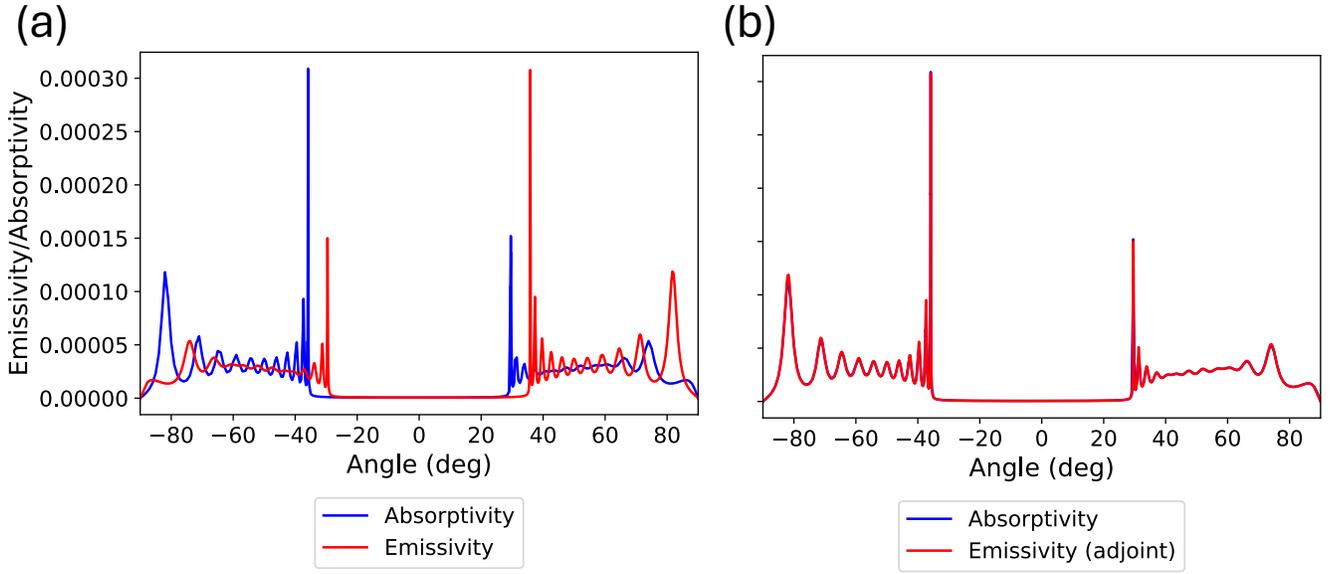

FIG. 1. (a) Emissivity and absorptivity of same nonreciprocal multilayer structure. (b) Emissivity of adjoint structure and absorptivity of nonreciprocal multilayer structure.

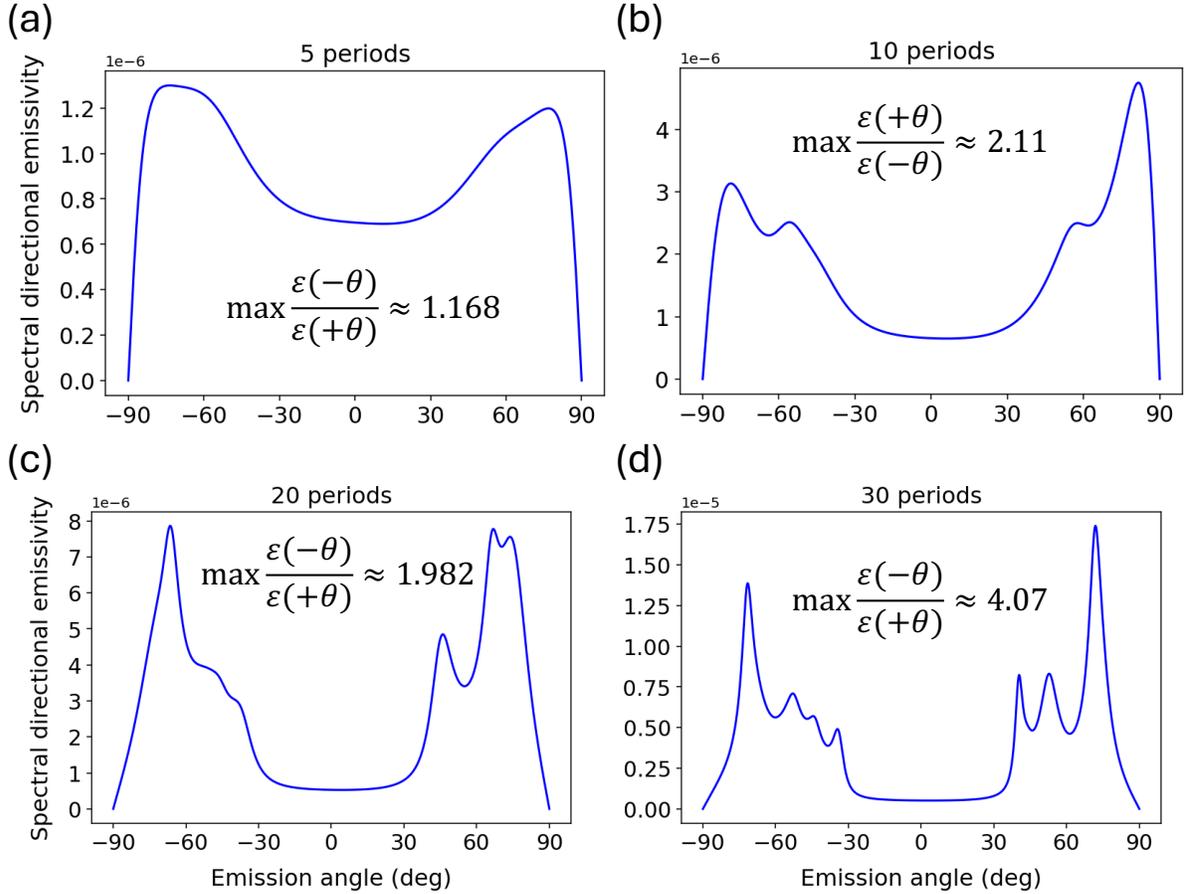

FIG. 2. Emissivity of the nonreciprocal multilayer structure with (a) 5 periods, (b) 10 periods, (c) 20 periods, and (d) 30 periods of our multilayer structure. The maximum emission contrast ratios are indicated for each structure.

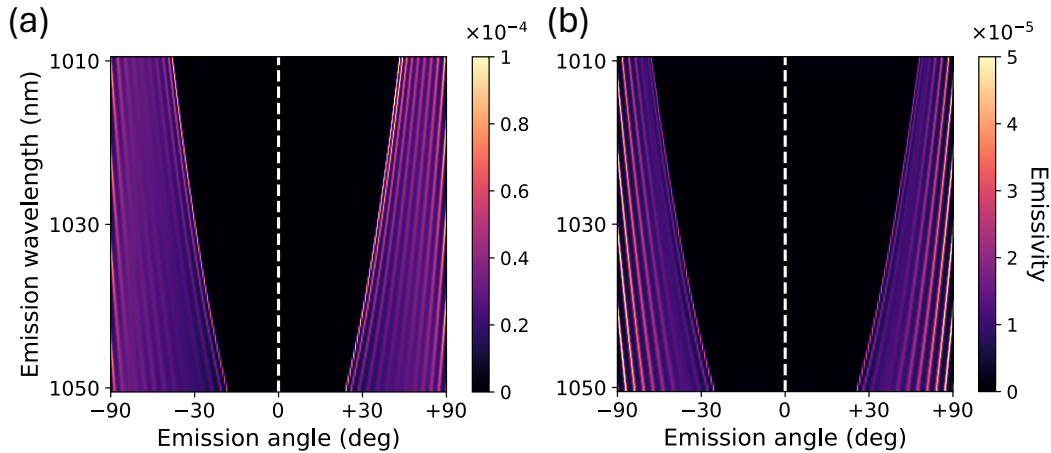

FIG. 3. Emissivity of the nonreciprocal multilayer structure under an applied external **B** field as a function of frequency and emission angle for the (a) TM (b) TE polarizations.

We further optimized for the separation between the band gap edge on the left hand side of the angular plot ($-\theta$ values):

$$\min_{\mathbf{t} \in \mathbb{R}^3} \frac{1}{\max A(-\theta)/A(+\theta)} + \frac{1}{|\theta_{A^s,BG} - \theta_{A,BG}|} \tag{21}$$

$$\text{s.t.} \quad 0 < |\theta| < 30° \tag{22}$$

where $|\theta_{A,BG}|$ is the angle (on the $-\theta$ side) corresponding to where the bandgap starts. In other words, we aimed to maximize the difference $|\theta^- - \theta^+|$, using the notation from the main text.

## V. PLOTS OF EMISSIVITY FOR EACH POLARIZATION

In Fig. 3, we provide the 2D emissivity plots as a function of frequency and emission angle for the TM and TE polarizations. We see that the asymmetry with respect to $\theta = 0$ is present in the TM polarization emissivity due to nonreciprocity, whereas the TE emissivity remains symmetric.

## VI. $S(\omega)$ FUNCTION

The $S(\omega)$ function used to compute Fig. 3a in the main text was obtained by fitting to the emission cross section of Yb:YAG [5, 6]. The function was then normalized to have integrated area of 1 over the bandwidth $\lambda = 1010-1050$nm.

## VII. MASS ATTENUATION OF SCINTILLATOR

In this section, we show that a uniform distribution of emitters along the depth of our $21.4\mu$m multilayer structure is a reasonable assumption. We use the mass attenuation data for YAG given in Ref. [7] and density value of $4.6$g/cm$^3$ [8].

For a photon energy of $10^{-1}$ MeV, the mass attenuation coefficient is $\approx 0.477$ cm$^2$/g. Using the Beer-Lambert law, we can compute the intensity of the x-ray beam after irradiating our structure:

$$I(z = 21.4\text{e-4 cm}) = I_0 e^{-0.477*4.6*21.4} \approx 0.995 I_0 \tag{23}$$

where $I_0$ is the incident x-ray intensity.

At higher photon energies such as $10^2$ MeV, the mass attenuation coefficient is $\approx 0.034$ cm$^2$/g, and we have:

$$I(z = 21.4\text{e-4 cm}) = I_0 e^{-0.034*4.6*21.4} \approx 0.9997 I_0 \tag{24}$$

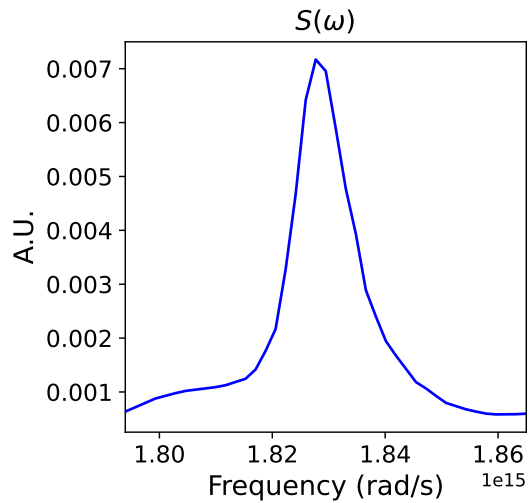

FIG. 4. $S(\omega)$ function interpolated from emission cross section data for scintillator Yb:YAG. The integrated area is normalized to 1.

Thus, our assumption is valid over the photon energy range of $10^{-1} - 10^2$ MeV for our structure.

However, for lower photon energies, such as $10^{-2}$ MeV, the mass attenuation coefficient is $\approx 39.63$ cm$^2$/g and we have:

$$I(z = 21.4\text{e-4 cm}) = I_0 e^{-39.63*4.6*21.4} \approx 0.677 I_0 \qquad (25)$$

In this low-energy photon regime, the uniform emission assumption would no longer hold.

From another perspective, the difference between assuming a uniform distribution of emitters and an attenuated distribution is $< 1\%$ if we have:

$$I(z = d) = I_0 e^{-(\mu/\rho)*4.6*d} > 0.99 I_0 \qquad (26)$$

$$-(\mu/\rho)*4.6*d > \ln(0.99) \approx 0.01 \qquad (27)$$

$$(\mu/\rho)*d < 0.0022 \quad \text{cm}^3/\text{g} \qquad (28)$$

where $\mu/\rho$ is the mass attenuation coefficient and $d$ is the thickness of the structure.

---


\fi

\end{document}